\documentclass{birkjour_t2}

\usepackage[english]{babel}
\usepackage[utf8]{inputenc}

\usepackage{amsmath,comment}
\usepackage{graphics}
\usepackage{graphicx,color}
\usepackage{amsbsy,amssymb,amsmath,ulem}
\usepackage[all]{xy}

\usepackage{hyperref}

\usepackage{tikz}
\usetikzlibrary{positioning}

\newcommand{\JJ}{{\boldmath \mbox{$J$}}}

\newcommand{\XX}{{\boldmath \mbox{$X$}}}

\newcommand{\rr}{{\boldmath \mbox{$r$}}}

\newcommand{\QQ}{{\boldmath \mbox{$Q$}}}

\newcommand{\dr}{\mathrm{d}\mathbf{r}}

\begin{document}

\title{Entropy and Entropy Production in Multiscale Dynamics}

\author{Miroslav Grmela}
\address{\'{E}cole Polytechnique de Montr\'{e}al, C.P.6079 suc. Centre-ville, Montr\'{e}al, H3C 3A7,  Qu\'{e}bec, Canada}
\email{miroslav.grmela@polymtl.ca}

\author{Michal Pavelka}
\address{Mathematical Institute, Faculty of Mathematics and Physics, Charles University, Sokolovsk\'{a} 83, 186 75 Prague, Czech Republic}

\author{V{\' a}clav Klika}
\address{Czech Technical University, Department of Mathematics – FNSPE, Trojanova 13, 120 00 Prague, Czech Republic}

\author{Bing-Yang Cao}
\address{Department of Engineering Mechanics, Tsinghua University, Beijing, 100084 China}

\author{Nie Bendian}
\address{Department of Engineering Mechanics, Tsinghua University, Beijing, 100084 China}



 \date{}

\maketitle

\begin{abstract}

Heat conduction is investigated on three levels: equilibrium, Fourier, and Cattaneo. The Fourier level is either  the point of departure  for investigating the approach to equilibrium or the final stage in the investigation of the approach from the Cattaneo level. Both investigations bring to the Fourier level  an entropy and a thermodynamics.   In the absence of external and internal influences preventing the approach to equilibrium the entropy that arises in the latter investigation is the production of the classical entropy that arises in the former investigation. If the approach to equilibrium is prevented, then the entropy that arises in the investigation of the approach from the Cattaneo level to the Fourier level still brings to the Fourier level the entropy and the thermodynamics even if the classical entropy and the classical thermodynamics is absent. We also note that vanishing total entropy production as a characterization of equilibrium state is insufficient.

\end{abstract}

\section{Introduction}

Macroscopic systems that are free from external forces and from  external and internal constraints reach states, called equilibrium states, at which their behavior is found to be well described by the classical equilibrium thermodynamics (ET). This is the experimental observation on which ET stands.  For one component macroscopic systems, the variables parametrizing the equilibrium states are: the volume $V$, the number of moles $N$, and the energy $E$.
A more detailed  investigation of the time evolution bringing the macroscopic systems to the equilibrium states (i.e. a more detailed investigation of the process of preparation of macroscopic systems for ET) reveals that the time evolution describing it  is driven by a potential. This potential, if evaluated at the asymptotically  reached equilibrium states, becomes the equilibrium entropy $S^{(ET)}$. The preparation process thus plays two roles: (i) it brings the macroscopic systems to equilibrium states where ET is applicable, (ii)  and it  also determines the fundamental thermodynamic relation $S^{(ET)}=S^{(ET)}(V,N,E)$ in which the individual nature of macroscopic systems is expressed in ET.
Formally, we represent the process of preparing macroscopic systems to ET  by the diagram
\begin{equation}\label{MET}
 \begin{tikzpicture}
  \node(M){M};
  \node(ET)[below of = M]{ET};
  \path[->]
	  (M) edge (ET);
 \end{tikzpicture}
\end{equation}
where $M$ represents the mesoscopic theory on which the preparation process is observed. The dynamics  involved in (\ref{MET}) will be called \textit{reducing dynamics}.

Let us now consider two  mesoscopic theories: $M$ and $m$. Both are assumed to be well established and the mesoscopic theory $m$ is more macroscopic than $M$. We say that $m$ is more macroscopic (or equivalently less microscopic) than $M$ if  some details seen in $M$ are not seen in $m$. We say that a mesoscopic theory is well established if its consequences agree with experimental observations from which the theory sprang out. For example,  we can think of $M$ as being the kinetic theory and $m$ as fluid mechanics. Note that the experimental observations on which $M$ is based are different from those on which $m$ is based. Since both $M$ and $m$ are well established, it must be  possible to prepare the macroscopic systems for the mesoscopic theory $m$. The preparation process (the reducing dynamics)
\begin{equation}\label{Mm}
 \begin{tikzpicture}
  \node(M){M};
  \node(m)[below of = M]{m};
  \path[->]
	  (M) edge (m);
 \end{tikzpicture}
\end{equation}
has to be seen in $M$. If the reducing dynamics (\ref{Mm}) is driven by a potential (similarly as the reducing dynamics (\ref{MET}) is)  then we can suggest to interpret the potential as an $(M\rightarrow m)$-entropy. If evaluated at the asymptotically reached state it becomes a fundamental thermodynamic relation in $m$ determined by (\ref{Mm}).

Next, we consider three well established theories $M,m$ and $ET$ and investigate relations depicted in the diagram
\begin{equation}\label{MmET}
 \begin{tikzpicture}
  \node(M){M};
  \node(m)[below of = M]{m};
  \node(ET)[below of = m]{ET};
  \path[->]
	  (M) edge [bend right=60]  (ET)
	  (m) edge (ET)
	  (M) edge (m);
 \end{tikzpicture}
\end{equation}
In the illustration that is  worked  out below  we consider  $ET$ to be the equilibrium,  $M$  the Cattaneo,  and $m$ the Fourier theories  of heat conduction.

\section{Equilibrium theory: ET}

We limit ourselves to  processes in which the volume $V$ and the number of moles $N$ remain unchanged. Hereafter, we therefore omit $V$ and $N$ and consider only the energy $E$ as the state variable in ET. The fundamental thermodynamic relation in ET is
\begin{equation}\label{fundET}
S^{(ET)}=S^{(ET)}(E)
\end{equation}
We call hereafter the quantity $S^{(ET)}$ an ET-entropy and the function $S^{(ET)}(E)$ an ``$ET$'' a fundamental thermodynamic relation.

We also introduce an ``$ET$'' thermodynamic potential
\begin{equation}
	\Phi^{(ET)}(E,E^*)\\=-S^{(ET)}(E)+E^*E. 
\end{equation}
The time evolution describing the  process occurring in the contact with a thermal bath with the inverse temperature $E^*$ is generated by
\begin{equation}\label{ETevol}
\dot{E}=-\Lambda^{(ET)}\Phi^{(ET)}_E
\end{equation}
where the dot  denotes the time derivative, $\Phi^{(ET)}_E=\frac{\partial \Phi^{(ET)}}{\partial E}$, and $\Lambda^{(ET)}>0$ is a parameter. The Lyapunov theorem ($\Phi^{(ET)}$ serves as the Lyapunov function) implies that $E\rightarrow E^{(ET)}$ as $t\rightarrow\infty$, where $E^{(ET)}(E^*)$ is the energy $E$ at which $\Phi^{(ET)}$ reaches its minimum, i.e. $E^{(ET)}(E^*)$ is a solution of $\Phi^{(ET)}_E=0$. The new entropy after the equilibration process with the thermal bath is completed is the Legendre transformation $S^{*(ET)}(E^*)= \Phi^{(ET)}(E^{(ET)}(E^*),E^*)$ of the entropy $S^{(ET)}(E)$.

Note that this level of description, referred to as equilibrium theory, is the only level in this article where the system is not closed (and interactions with the surroundings is considered). As a result, energy of the system is not conserved. The choice of the evolution equation \eqref{ETevol} can be understood via the Lyapunov theorem as the system is guaranteed\footnote{Lyapunov theorem rigorously applies only to dynamical systems of finite dimension and further it needs to be shown that $\Phi^{(ET)}(E^{(ET)})=0$ and $\Phi^{(ET)}(E)>0$ otherwise. The former requirement corresponds to the above observation that the equilibrium energy $E^{(ET)}$is the Legendre transformation of the entropy $S^{(ET)}(E)$. The latter condition represents the observation that the fundamental thermodynamic potential is extremal in equilibrium. } to evolve towards the equilibrium characterised by the minimal energy $E^{(ET)}$.

\section{Fourier theory: m $\rightarrow$ ET}\label{Ftheory}

Now we turn to the  Fourier theory ``$m$''.  There is only one  state variable in the Fourier theory. It is the field of the internal energy $e(\rr)$;  $\rr\in\mathbb{R}^3$ is the position vector. The state space in $m$ will be denoted by $U^{(m)}$ (i.e. $e(\rr)\in U^{(m)}$).

The time evolution of $e(\rr)$  (i.e. the time evolution  in the reducing dynamics (\ref{MET}) in which the level $M$ is replaced by the level $m$)  is governed by
\begin{equation}\label{Four1}
\frac{\partial e}{\partial t}=-\nabla\cdot(\Lambda^{(m)} \nabla e^*)
\end{equation}
We explain the meaning of the symbols introduced in (\ref{Four1}).

By $e^*(\rr)$  we denote a state variable that is conjugate to $e(\rr)$. We define it as follows. We introduce first
\begin{equation}\label{fundm}
s^{(m)}:U^{(m)}\rightarrow \mathbb{R}
\end{equation}
called a m-entropy.
We assume that $s^{(m)}$
is  a sufficiently regular and concave function. We shall call $s^{(m)}(\rr)= s^{(m)}(e;\rr)$  introduced in (\ref{fundm}) a ``$m$'' thermodynamic relation similarly as we call  (\ref{fundET}) $S^{(ET)}=S^{(ET)}(E)$  a ``$ET$'' fundamental thermodynamic relation.   The conjugate state variable $e^*(\rr)$ is  introduced as $e^*(\rr)=s^{(m)}_{e(\rr)}$, where we use the notation $e^*(\rr)=s^{(m)}_{e(\rr)}=\frac{\delta s^{(m)}}{\delta e(\rr)}$, with $\delta.../\delta...$ an appropriate functional derivative. The symbol $\nabla$ stands for $\partial/\partial \rr$. By $\Lambda^{(m)}(e(\rr))$ we denote a positive definite operator.
Hereafter, we shall use the summation convention: $\nabla\cdot(\Lambda^{(m)} \nabla e^*)=\partial_i(\Lambda^{(m)}_{ik}\partial_ke^*)=\sum_{i=1}^3\sum_{k=1}^3\partial_i(\Lambda^{(m)}_{ik}\partial_ke^*)$, where $\partial_i= \partial/\partial r_i; i=1,2,3$

\subsection{Properties of solutions to Eq.(\ref{Four1})}\label{2.0}

We make a few observations about solutions to (\ref{Four1}). First, we note that
\begin{equation}\label{Edot}
\dot{E}=0
\end{equation}
(where  dot means the time derivative and  $E=\int d\rr e(\rr)$)   provided the boundary conditions are chosen (in accordance with  the assumption of the absence of external influences) in such a way that the integrals over the boundary equal zero. Therefore the ET level corresponding to the reduction of the Fourier $m$ level that is compliant with the Fourier evolution equation \eqref{Four1} is the fixed equilibrium $E^{(ET)}$ with the fundamental thermodynamic relation $S^{(ET)}=S^{(ET)}(E^{(ET)})$.

The second observation is about the time evolution of the m-level entropy $s^{(m)}$. We note that
\begin{equation}\label{Smdot}
\dot{S}^{(m)}>0
\end{equation}
where $S^{(m)}=\int d\rr s^{(m)}(e;\rr)$. Indeed, $\dot{S}^{(m)}= -\int d\rr e^*\nabla\cdot(\Lambda^{(m)}\nabla e^*)=\int d\rr (\nabla\epsilon^*)\cdot\Lambda^{(m)} (\nabla e^*)\geq 0$ due to the assumption that $\Lambda^{(m)}$ is a  positive definite operator. The entropy production is thus being equal to $\dot{S}^{(m)}$ as there is no entropy flux at the boundary assumed, and on the m-level it is thus $\int d\rr (\nabla\epsilon^*)\cdot\Lambda^{(m)} (\nabla e^*)>0$.

With an introduction of a ``m''-thermodynamic potential
\begin{equation*}
  \Phi^{(m)}(E^*)=\int d\rr \phi^{(m)}(\rr; E^*)
\end{equation*}
where 
\begin{equation}\label{phimdef}
\phi^{(m)}(\rr;E^*)=-s^{(m)}(e;\rr)+E^* e(\rr)
\end{equation}
the above two observations  (\ref{Edot}) and (\ref{Smdot}) then imply
\begin{equation}\label{Phimdot}
\dot{\Phi}^{(m)}<0.
\end{equation}

(i) \textit{The approach to equilibrium}

The asymptotic , $t\rightarrow \infty$,  solutions to  (\ref{Four1}), denoted $e^{(ET)}(\rr;E^*)$,  are minima of $\Phi^{(m)}$ (i.e. solutions to $\Phi^{(m)}_{e(\rr)}=0$ ). Indeed, the m-level thermodynamic potential $\Phi^{(m)}$ plays the role of the Lyapunov function for the $t\rightarrow \infty$ approach to the equilibrium states  $e^{(ET)}(\rr;E^*)$ \footnote{Lyapunov theory per se is not available in this generality. Although the problem of stability in (nonlinear) partial differential equations is more complex and requires tailored analysis to a given problem, we propose to consider the existence of Lyapunov type functional as a strong indication of stability of equilibrium point. This is supported by extensions of Lyapunov theory to certain classes of partial differential equations \cite{Buis, Temam}}.

(ii) \textit{The ``$ET$'' fundamental thermodynamic relation (\ref{fundET}) implied by the ``$m$'' fundamental thermodynamic relation (\ref{fundm})}


The approach to equilibrium $E^{(ET)}$ from level $m$ should be compliant with the ET description, i.e. $\Phi^{(m)}(E^*) = \Phi^{(ET)}(E^*)$, where $\Phi^{(ET)}(E^*)$ is the Legendre transformation of $S^{(ET)}(E)$ and similarly $\Phi^{(m)}(E^*)$ is the Legendre transformation of $s^{(m)}(e)$. Then
\begin{equation*}
  -S^{(ET)}(E)+E^* E|_{E=E^{(ET)}(E^*)} = \Phi^{(ET)}(E^*) = \Phi^{(m)}(E^*) = -\int \dr s^{(m)}(e^{(ET)}(E^*,\rr);\rr) + E^* E,
\end{equation*}
and we see that $E^*$ introduced in (\ref{phimdef} is the conjugate variable to $E$ on the level ET (i.e. $E^*=S_E$), equilibrium variables from the two levels satisfy the following correspondence $E^{(ET)}=E=\int d\rr e(\rr)|_{e=e^{(ET)}}=\int d\rr e^{(ET)}(\rr;E^*)$, and entropies are related as  $S^{(ET)}(E^*)=S^{(m)}(e^{(ET)}(\rr;E^*);\rr) = \int \dr s^{(m)}(e^{(ET)}(E^*,\rr);\rr)$.

We also note that that if we choose the m-entropy $s^{(m)}(e(\rr))$ to be pointwise (in energy $e$) the same function as $S^{(ET)}(E)$ (the so called local equilibrium assumption) then $S^{(ET)}=S^{(ET)}(E)$ is the ``$ET$'' fundamental thermodynamic relation implied by the ``$m$'' fundamental thermodynamic relation.

Note also that it is possible to construct a ``Lyapunov potential'' $\Phi(e) - \langle \Phi_e|_{e_0}, (e-e_0)\rangle$ which leads to a non-equilibrium steady state (energy density field $e_0$), see \cite{Vitek}.

Finally, we note that the Fourier equation (\ref{Four1}) can also be written in the form
\begin{equation}\label{Four3}
\frac{\partial e}{\partial t}=\nabla\cdot(\Lambda^{(m)} \nabla \Phi^{(m)}_e)
\end{equation}
with the ``$m$'' thermodynamic potential $\Phi^{(m)}$ given in (\ref{phimdef}).

\subsection{Generalized Fourier theory: $m\rightarrow ET$}\label{2.1}

Still another way to write the  Fourier time evolution equation (\ref{Four1}) is
\begin{equation}\label{Four2}
\frac{\partial e}{\partial t}=\Psi^{(m)}_{e^*(\rr)}
\end{equation}
where $\Psi^{(m)}=\frac{1}{2}\int d\rr (\nabla e^*)\cdot\Lambda^{(m)} (\nabla e^*)$. Indeed, $\Psi^{(m)}_{e^*(\rr)}=-\nabla\cdot(\Lambda^{(m)} \nabla e^*)$. We call $\Psi^{(m)}$ a ``$m$''  dissipation potential.

We  note that both (\ref{Edot}) and (\ref{Smdot}) and thus also (\ref{Phimdot}) remain to hold also for more general dissipation potentials $\Psi$. The  properties that guarantee (\ref{Edot}), (\ref{Smdot}),  (\ref{Phimdot})
are the following:
(i) $\Psi^{(m)}$ is a sufficiently regular function $\Psi^{(m)}:U^{(m)}\times U^{(X)} \rightarrow\mathbb{R}; \,\,(e(\rr),X^{(m)}(\rr))\mapsto \Psi^{(m)}(e,X)$,  with $X^{(m)}=\nabla e^*$,
 (ii) $\Psi^{(m)}(e(\rr),0)=0$; (iii) $\Psi^{(m)} $ as a function of $X^{(m)}(\rr)\in U^{(X)}$ reaches its minimum at $0$, and (iv)  $\Psi^{(m)}$ as a function of $X^{(m)}(\rr)\in U^{(X)}$ is a convex function in a neighborhood of  $0$.

We indeed easily verify that (\ref{Edot}) and (\ref{Smdot})  hold and that \\$\Psi^{(m)}=\frac{1}{2}\int d\rr (\nabla e^*)\cdot\Lambda^{(m)} (\nabla e^*)$ is a particular case of the dissipation potential satisfying the four properties of $\Psi^{(m)}$ listed above. The quantity $X^{(m)}$ is called a ``$m$'' dissipative thermodynamic force or just simply a ``$m$'' thermodynamic force. We note that  $X^{(m)}=\nabla e^*$ can also be written as  $X^{(m)}=-\nabla \Phi^{(m)}_e$.

\section{Cattaneo theory: $ M  \rightarrow ET $}\label{Cattth}

In order to extend the range of applicability of the Fourier theory (e.g. to investigations of the   heat conduction in electronic devices)
we follow Cattaneo \cite{Cattaneo} and extend the state space of the Fourier theory $U^{(m)}$ to a larger state space $U^{(M)}$. The elements of $U^{(M)}$ are the fields $e(\rr)$, that serve as the state variables in the Fourier theory,  and an additional vector field  $\JJ(\rr)$:
\begin{equation}\label{Cattstv}
(e(\rr),\JJ(\rr))\in U^{(M)}
\end{equation}
The  physical interpretation of $\JJ(\rr)$ will be revealed later in this section in the investigation of the time evolution of (\ref{Cattstv}).

Similarly as in ET or in the Fourier theory, we introduce the M-entropy
\begin{equation}\label{fundM}
s^{(M)}:U^{(M)}\rightarrow \mathbb{R}
\end{equation}
and call $s^{(M)}(\rr)=s^{(M)}(e,\JJ;\rr)$ a ``$M$'' fundamental thermodynamic relation. Again, keeping the notation introduced in ET and in the Fourier theory,  we introduced the conjugate state variables $e^*=s^{(M)}_{e(\rr)}, \JJ^*=s^{(M)}_{\JJ(\rr)}$ and the ``$M$'' thermodynamic potential $\Phi^{(M)}=-S^{(M)}+E^*E$, where $S^{(M)}=\int d\rr s^{(M)}(e,\JJ;\rr)$, and $E=\int d\rr e(\rr)$.

In  the time evolution of the Cattaneo state variables (\ref{Cattstv}), we want first of all to preserve (\ref{Edot}) and (\ref{Smdot}) but with $S^{(M)}$ replacing $S^{(m)}$. It is easy to verify that in the time evolution governed by
\begin{equation}\label{Catt}
\frac{\partial}{\partial t}\left(\begin{array}{cc}e\\J_i\end{array}\right)=\left(\begin{array}{cc}\partial_i\left(\frac{1}{(e^*)^2}J^*_i\right)\\\frac{1}{(e^*)^2}\partial_i(e^*)\end{array}\right)
+\left(\begin{array}{cc}0\\\Psi^{(M)}_{J^*_i}\end{array}\right)
\end{equation}
both (\ref{Edot}) and (\ref{Smdot}) with $S^{(M)}$ replacing $S^{(m)}$ remain valid. By $\Psi^{(M)}$ we denote that ``$M$'' dissipation potential satisfying the  properties listed in Section \ref{2.1} with the ``$M$''  thermodynamic force $X^{(M)}=\JJ^*$.  We shall discuss the properties of solutions to (\ref{Catt}) below in Section \ref{3.1} and derive (\ref{Catt}) in Section \ref{3.2}.

\subsection{Properties of solutions to Eq.(\ref{Catt})}\label{3.1}

The energy conservation (\ref{Edot}) of the energy $E=\int d\rr e(\rr)$ is manifestly visible in (\ref{Catt}). The ``$M$''  energy flux (i.e. the  heat flux on the M-level denoted by the symbol $\mathfrak{Q}^{(M)}$) is related to the vector field $\JJ$ (that serves as the extra state variable) by
\begin{equation}\label{heatM}
\mathfrak{Q}^{(M)}=-\frac{1}{(e^*)^2}\JJ^*
\end{equation}

Now we turn to the entropy inequality (\ref{Smdot}) with $s^{(m)}$ replaced by $s^{(M)}$. We see immediately  that
 $\frac{\partial s^{(M)}}{\partial t}=\partial_i\left(\frac{1}{e^*}J_i^*\right)+J_i^*\Psi^{(M)}_{J_i^*}$. The ``$M$'' entropy flux,  denoted by the symbol $\mathfrak{S}^{(M)}$, is thus given by
\begin{equation}\label{sflM}
\mathfrak{S}^{(M)}=-\frac{1}{e^*}\JJ^*
\end{equation}
From (\ref{heatM}) and (\ref{sflM}) we then see that $\mathfrak{S}^{(M)}=e^*\mathfrak{Q}^{(M)}$ which is indeed the classical relation between the heat flux and the entropy flux.

The ``$M$'' entropy production  implied by (\ref{Catt}) is given by
\begin{equation}\label{sigmaM}
J_i^*\Psi^{(M)}_{J_i^*}>0
\end{equation}
The inequality sign in (\ref{sigmaM}) is a direct consequence of  the four  properties of dissipation potentials listed in Section \ref{2.1}. Since both $\dot{E}=0$ and $\dot{S}^{(M)}>0$ hold  then also the inequality $\dot{\Phi}^{(M)}<0$ holds. This inequality then implies (see more in Section \ref{rigor}) the approach to equilibrium and the ``$ET$''  fundamental thermodynamic relation implied by the ``$M$''  fundamental thermodynamic relation (\ref{fundM}). 
We note that the ``$ET$'' fundamental thermodynamic relations obtained in Section \ref{2.0} and in this section are identical if $\JJ^{(ET)}=0$ and $[s^{(M)}]_{\JJ=0}=s^{(m)}$.

In order that the Cattaneo equation (\ref{Catt}) be regarded as an extension of the Fourier equation (\ref{Four1}) or (\ref{Four2}), we have to show that solutions to (\ref{Four1}) approximate well asymptotic solutions to (\ref{Catt}). If this were the case then the macroscopic system under investigation can be prepared (by letting the time evolution to take its course  for a sufficiently long time) for the m-level description. We shall investigate this question in Section \ref{5}.

Now we return to the question of what is the physical interpretation of the vector field $\JJ(\rr)$ that serves on the level ``$M$'' as the extra state variable. We see from
(\ref{heatM}) and (\ref{sflM}) that $\JJ(\rr)$ is related to but it is not the same as either heat flux or the entropy flux. The relation involves the fundamental thermodynamic relation (\ref{fundM}). Another information about the physical interpretation of $\JJ(\rr)$ will arise in Section \ref{3.2} where we discuss derivation of (\ref{Catt}).

\subsubsection{Rigorous derivation of the approach to ET: Open problem}\label{rigor}

In the context of the Fourier equations (\ref{Four1}) and (\ref{Four2}) (considered together with the boundary conditions expressing the absence of external  forces),  the entropy production disappears (i.e. $\dot{\Phi}^{(m)}=-\dot{S}^{(m)}=0$ due to energy conservation $\dot{E}^{(m)}=0$ and as entropy production corresponds to the whole time derivative of entropy $S^{(m)}$ due to zero entropy flux) only at the equilibrium sates (i.e. the states at which the thermodynamic potential $\Phi^{(m)}$ reaches its minimum). This then makes a strong indication of the approach to equilibrium via Lyapunov theorem as discussed above. The situation is different in the context of the Cattaneo equation (\ref{Catt}). The entropy production disappears (i.e. $\dot{\Phi}^{(M)}=-\dot{S}^{(M)}=0$ again due to conservation of total energy and because the system is assumed closed resulting in zero total entropy flux; finally note that a total entropy production rather than local is considered) on the manifold $\{(e,\JJ)\in U^{(M)}|\JJ^*=0\} $ while the equilibrium states form a smaller submanifold $\{(e,\JJ)\in U^{(M)}|\JJ^*=0, e^*=E^*\} $. It is the mutual interaction of the dissipative (governed by the second term on the right hand side of (\ref{Catt})) and the nondissipative (governed by the first term on the right hand side of (\ref{Catt})) time evolutions  that is expected  do drive solutions to the Cattaneo equation  to the equilibrium states. The similar situation arises in the context of the Boltzmann kinetic equation where the entropy production disappears at the local Maxwell distribution functions and the equilibrium states are the total Maxwell distribution functions that is a small submanifold of the manifold formed by the local Maxwell distribution functions. An interesting open problem is to adapt the rigorous proof \cite{Vill1} of the approach to equilibrium for the Boltzmann equation  to the Cattaneo equation (\ref{Catt}).

\subsection{Derivation of the Cattaneo equation (\ref{Catt})}\label{3.2}

Both the Fourier  (\ref{Four1}) and the Cattaneo (\ref{Catt}) time evolution equations have been derived in the previous sections by investigating their consequences. We have first proposed the equations and then we have shown that they both generate the time evolution describing the approach to equilibrium at which the classical equilibrium thermodynamics applies. In addition, in Section \ref{5},  we shall also show  that solutions to (\ref{Catt}) agree with solutions to  (\ref{Four1}) provided the parameters entering  (\ref{Catt}) and the initial condition are chosen appropriately. Now, we turn to the derivation of (\ref{Catt}) that begins with some fundamental principles and Eq.(\ref{Catt})  arises  from an  investigation of their consequences.
The  fundamental principles can be found either  in the very microscopic (atomistic) viewpoint of heat or  in some general considerations  about the mathematical structure of mesoscopic theories. We shall now  derive (\ref{Catt}) from the principles that have arisen on the latter route.

As argued in \cite{GO},\cite{OG}, \cite{O},\cite{PKG}, the mesoscopic time evolution equations describing the approach to equilibrium have all a general structure called GENERIC. The vector field generating the mesoscopic time evolution (i.e. the right hand side of the time evolution equation)  is a sum of two parts, one (being a remnant of the Hamiltonian dynamics of the fundamental particles composing the macroscopic system under investigation) is Hamiltonian, and the other (driving the system to the thermodynamic equilibrium) is gradient. We now proceed to recognize the GENERIC structure  in (\ref{Catt}).

The second term on the right hand side of (\ref{Catt}) represents a general gradient dynamics. The requirement of the GENERIC structure thus does not bring anything new to the second term on the right hand side of (\ref{Catt}). The different situation is however with the first term on the right hand side of (\ref{Catt}). According to GENERIC, this term has to be Hamiltonian with the energy $E=\int d\rr e(\rr)$ serving as the generating potential and the entropy $S^{(M)}=\int d\rr s^{(M)}(\rr)$ playing the role of  the Casimir potential (i.e. a potential that is different from the energy but, as the energy, remains unchanged during the Hamiltonian time evolution).  We recall that the Hamiltonian vector field is a covector field (that is the gradient of the energy $E$)  transformed into a vector field  by a Poisson bivector $\mathcal{L}$. The Poisson bivector is then expressed mathematically in the Poisson bracket $\{A,B\}=\int d\rr (A_e,A_{\JJ})^T\mathcal{L}(B_e,B_{\JJ})$, where $A$ and $B$ are real valued and sufficiently regular functions of $(e(\rr),\JJ(\rr))$, and the vector $(A_e,A_{\JJ})^T$ is the transpose of the vector $(A_e,A_{\JJ})$. A bivector $\mathcal{L}$ is a Poisson bivector if the  bracket $\{A,B\}$ is the Poisson bracket (i.e. $\{A,B\}=-\{B,A\}$ and the Jacobi identity $\{A,\{B,C\}\}+\{B,\{C,A\}\}+\{C,\{A,B\}\}=0$ holds). With the Poisson bracket $\{A,B\}$ we can write the Hamiltonian time evolution as $\dot{A}=\{A,E\}, \forall A$ and the Casimir potential $C$ as the potential for  which $\{A,C\}=0,\forall A$.

We now proceed  to show that the first term on the right hand side of (\ref{Catt}) is Hamiltonian.
To construct a Hamiltonian vector field we need a potential (that has the physical interpretation of the energy $E$) and a Poisson bivector $\mathcal{L}$ (that has the physical interpretation  of  kinematics). Having chosen the state variables $(e(\rr),\JJ(\rr))$, the energy is given by $E=\int d\rr e(\rr)$ (i.e. the gradient of the energy is $(E_{e(\rr)},E_{\JJ(\rr)})=(1,0)$). It remains thus to identify  the Poisson bivector $\mathcal{L}$ expressing kinematics of the state variables $(e(\rr),\JJ(\rr))$. To find it, we turn to  the physical origin of heat.

We begin with seeing  the heat as a gas of phonons. Since we require that the entropy $S^{(M)}$ is the Casimir, it is useful to start  the search for kinematics  of the   state variables $(s^{(M)}(\rr),\JJ(\rr))$ rather than the state variables $(e(\rr),\JJ(\rr))$. The relation between $(e(\rr),\JJ(\rr))$ and  $(s^{(M)}(\rr),\JJ(\rr))$ is given by $\JJ(\rr)=\JJ(\rr)$ and the ``$M$'' fundamental thermodynamic relation (\ref{fundM}).
We shall  assume that the relation between  $(e(\rr),\JJ(\rr))$ and  $(s^{(M)}(\rr),\JJ(\rr))$ is one-to-one.  This assumption  restricts the choice of the ``$M$'' fundamental thermodynamic relations. The restriction is in fact a weak version of the local equilibrium assumption. Indeed,
if we interpret $e^*(\rr)$ as an inverse of the local absolute temperature then  $e^*(\rr)>0$ and the transformation $(e(\rr),\JJ(\rr)) \rightleftarrows (s^{(M)}(\rr),\JJ(\rr))$ is  one-to-one.

The Poisson bracket expressing kinematics of  $ (s^{(M)}(\rr),\JJ(\rr))$, where the field $\JJ(\rr)$ has the physical interpretation of the phonon momentum divided by $s^{(M)}$, is given by
\begin{equation}\label{PBph}
\{A,B\} = \int d\rr (\partial_k A_{s^{(M)}} B_{J_k} - \partial_k B_{s^{(M)}} A_{J_k}) + \int d\rr \frac{1}{s^{(M)}}(\partial_i J_j -\partial_j J_i)A_{J_i} B_{J_j}	
\end{equation}
This has been
shown in
 Section 3.9 of  \cite{PKG}). We then directly verify that the  bracket obtained by transforming the first term on the right hand side of (\ref{PBph}) from the state variables  $(s^{(M)}(\rr),\JJ(\rr))$ to the state variables $(e(\rr),\JJ(\rr))$ (we recall that we are assuming that the transformation is one-to-one) implies (by using  $\dot{A}=\{A,E\}, \forall A$) the time evolution governed by (\ref{Catt}) (without the second term on its right hand side). As for the second term on the right hand side of (\ref{PBph}), we shall show in Section \ref{5} below that if we limit ourselves to states  that are not too far from states at which the time evolution is governed by the Fourier equation (\ref{Four1}) then the second term  is negligible, which is also the setting of the Symmetric Hyperbolic Thermodynamically Compatible (SHTC) equations \cite{ADER1}, \cite{SHTC-GENERIC}, developed by the Godunov school of thermodynamics.

 Summing up, we have shown that (\ref{Catt}) possesses GENERIC structure, with the vector field $\JJ(\rr)$ having the physical interpretation of the phonon momentum divided by entropy, provided we limit our investigation to the stage in the time evolution in which solutions to the Fourier equation (\ref{Four1}) provide a  good approximation to solutions to the Cattaneo equation  (\ref{Catt}).

An alternative investigation of the kinematics of   $(e(\rr),\JJ(\rr))$  can be based on the analogy between the time evolution of the mass density and the mass flux  and the time evolution of the energy density and the energy flux. This analogy  has already been noted and exploited in three different and apparently independent investigations. In \cite{GT}, it was an attempt to develop a Lagrangian formulation of the Cattaneo hydrodynamics, in \cite{Guo1},  it was the formulation of the thermo-mass viewpoint of heat, and in \cite{RA},  it was an investigation of the Grad hierarchy (the distinction made between the material F-fields and the caloric G-fields). In this paper we  mention only some  results. Instead of starting with the Poisson bracket (\ref{PBph}), we start with the standard mass-momentum Poisson bracket \cite{Arnold1}, \cite{PKG}
\begin{eqnarray}\label{PB1}
	\{A,B\}&=&\int d\rr\left[\hat{J}_i\left(\partial_j(A_{\hat{J}_i})B_{\hat{J}_j}-\partial_j(B_{\hat{J}_i})A_{\hat{J}_j}\right)\right.\nonumber\\
	&&\left. + s^{(M)}\left(\partial_i(A_{s^{(M)}})B_{\hat{J}_i}-\partial_i(B_{s^{(M)}})A_{\hat{J}_i}\right)\right]
\end{eqnarray}
where the field $s^{(M)}$ replaces the mass density and $\hat{\JJ}$ the momentum density related to the above vector field $\JJ$ through $\hat{\JJ} = s^{(M)} \JJ$. Brackets \eqref{PB1} and \eqref{PBph} are equivalent (being transformation of each other).

\section{M  $\rightarrow$ m}\label{4}

The Fourier theory represented by (\ref{Four1}), (\ref{Four3}),  and (\ref{Four2}) addresses systems that are allowed to approach equilibrium. We can however also include into the Fourier theory  systems that, due to the presence of  external and internal constraints,  are prevented from approaching equilibrium and thus from preparing them for ET level of description. We shall denote the external and internal influences preventing the approach to ET by the symbol $\mathcal{P}$.  If we still limit ourselves only to the processes that preserve the energy, the equation replacing (\ref{Four1}) and  (\ref{Four3}) will take the form
\begin{equation}\label{Four4}
\frac{\partial e}{\partial t}=-\partial_i(\mathcal{J}_i)
\end{equation}
where the energy flux $\mathcal{J}$ remains unspecified. Its specification (as a function of $e(\rr)$ and the external and internal constraints $\mathcal{P}$)  will be called (in accordance with the established terminology) \textit{Constitutive Relation} or in an  abbreviated form CR.

We now assume that a comparison of results of experimental observations with solutions to Eq.(\ref{Four4}) (that is equipped  with an appropriate Constitutive Relation $\mathcal{J}=\mathcal{J}^{CR}(e,\mathcal{P})$  shows an agreement. This then means that any other, more microscopic (i.e. more detailed) and well established,  description has to show an approach to (\ref{Four4}).

The next question is of which more microscopic theory (more microscopic level $M$) we choose. In the illustration discussed in this paper
 we choose a theory in which a flux $\Upsilon(\rr)$ plays the role of the state variable. The relation between $\mathcal{J}$ and $\Upsilon$ will be clarified later.

 The simplest time evolution equation for $\Upsilon$ describing the approach to $\mathcal{J}^{CR}$ as $t\rightarrow\infty$ is then a direct adaptation of the Fourier equation (\ref{Four3}) to the  state variable $\Upsilon$:
\begin{equation}\label{Ups}
\frac{\partial\Upsilon_i}{\partial t}=-\Lambda^{(Mm)}_{ik}\Phi^{(Mm)}_{\Upsilon_i}
\end{equation}
where
\begin{equation}\label{UpsPhi}
\Phi^{(Mm)}(\Upsilon; \mathcal{J}^{CR})=-S^{(Mm)}(\Upsilon)+\mathcal{J}^{CR}_j\Upsilon_j
\end{equation}
is an ``$Mm$'' thermodynamic potential,  $\Lambda^{(Mm)}$ is a positive definite operator, and $S^{(Mm)}$ is the entropy associated with the $M\rightarrow m$ time evolution  (we call it, in accordance with the terminology used in the previous sections, an ``$Mm$'' entropy).
Indeed, $\dot{\Phi}^{(Mm)}=-\int d\rr \Phi^{(Mm)}_{\Upsilon_i}\Lambda^{(Mm)}_{ik}\Phi^{(Mm)}_{\Upsilon_k}<0$, and the Lyapunov theory suggests  the approach to $ \mathcal{J}^{CR}$, the assumed equilibrium of evolution \eqref{Ups}, as $t\rightarrow\infty$.

In order to emphasize the difference between the Fourier and the Cattaneo theories discussed in the previous sections (all addressing the approach to ET), we use  the symbols $\mathcal{J}$ and $\Upsilon$ instead of $\JJ$ to denote the fluxes and ``$\dag$'' instead of ``$*$'' to denote the conjugate variables (i.e. $\Upsilon^{\dag}=S^{(Mm)}_{\Upsilon})$.

It is advantageous to recapitulate dynamic reduction between levels of description \cite{PKG}. Dynamic reduction is used mainly to obtain MaxEnt value of direct variables while the conjugate variables are considered independent and their values corresponding to MaxEnt values of direct variables are not used. The reason is that the exact MaxEnt manifold corresponding to MaxEnt values of both direct and conjugate variables is not invariant to the reduced evolution because the reduced dynamics is typically dragging the evolution away from the MaxEnt manifold (the vector field is ``sticking out'' from the exact MaxEnt manifold). The suggested remedy of this issue is that the direct variables attain their MaxEnt values while the conjugate variables are corrected so that the reduced manifold is invariant to the vector field determining the reduced dynamics. This entails a change in the direct-conjugate variables relation from the reducing dynamics but which can be finally corrected by adaptation of entropy on the reduced level. Therefore in direct variables the evolution stays on the MaxEnt manifold and only in the extended contact geometry where both direct and conjugate variables are considered independent this is not so. The correction of entropy/energy then allows to return exactly to the MaxEnt manifold even in the contact geometry setting.

Since it is the conjugate $\Upsilon^{\dag}$ that approaches the Constitutive Relation $\mathcal{J}^{CR}(e,\mathcal{P})$ (direct variables attain their determined values during reduction, the MaxEnt values, while conjugate variables are used for assuring that the reduced evolution lives on the reduced MaxEnt manifold in direct variables), the energy flux $\mathcal{J}$ introduced in (\ref{Four4}) is related to the flux $\Upsilon$ introduced in (\ref{Ups}) by $\mathcal{J}=\Upsilon^{\dag}$.

The fundamental thermodynamic relation on the level ``$m$'' implied by the fundamental thermodynamic relation $S^{(Mm)}=S^{(Mm)}(\Upsilon)$ is $\Sigma^{(m)}=\Sigma^{(m)}(e,\mathcal{P})$, where
\begin{subequations}
\begin{eqnarray}\label{fundSig}
	\Sigma^{(m)}(e,\mathcal{P})&=&[S^{*(Mm)}(\Upsilon^{\dag})]_{\Upsilon^{\dag}=\mathcal{J}^{CR}(e,\mathcal{P})},\\
	S^{*(Mm)}(\Upsilon^{\dag})&=&\Phi^{(Mm)}(\Upsilon^{(m)}(e,\Upsilon^{\dag}),\Upsilon^{\dag}),
\end{eqnarray}
\end{subequations}
and $\Upsilon^{(m)}(e,\Upsilon^{\dag})$ is the state variable $\Upsilon$  approached  as $t\rightarrow\infty$, i.e the state variable $\Upsilon$ for which the ``$Mm$'' thermodynamic potential $\Phi^{(Mm)}$ reaches its minimum.  Hence $\Upsilon^{(m)}(e,\Upsilon^{\dag})$ is a solution to $\Phi^{(Mm)}_{\Upsilon}=0$. The symbol  $\Sigma^{(m)}$ was introduced to denote  $[S^{*(Mm)}]_{\Upsilon^{\dag}=\mathcal{J}^{CR}(e,\mathcal{P})}$  in order to simplify the notation and also in order to emphasize the difference between the two entropies
 $S^{(m)}(e)$ and $\Sigma^{(m)}(e,\mathcal{P})$. Additionally, the notation (starred entropies and fundamental thermodynamic potentials) is in accordance with the notation from earlier general discussion of reduction and extensions \cite{RedExt}. Both are  ``$m$'' level  entropies   but they have  very different origins. The former, the m-entropy $S^{(m)}$,  arises
in the analysis of the passage $m\rightarrow ET$ and the latter $\Sigma^{(m)}$, that we shall call a CRm-entropy,   in the analysis of the passage $M\rightarrow m$.  The former is a potential driving the $m\rightarrow ET$   time evolution, the latter is a  potential that arises  when the time evolution $M\rightarrow m$ has been completed.
In the presence of the external and internal influences $\mathcal{P}$ preventing the approach to ET only the CRm-entropy $\Sigma^{(m)}$ is present on the level ``$m$''. In the case when the approach to ET is allowed,  both the m-entropy $S^{(m)}$ and the CRm-entropy $\Sigma^{(m)}$. are present. Their relation is investigated in the next section.

\section{M $\rightarrow$ m $\rightarrow$ ET}\label{5}

In this section we consider systems that are not prevented to reach the level ``$ET$''. We also assume that comparison of results of experimental observations with solutions to both the Fourier and the Cattaneo theories shows an agreement. Our objective is to prove   that in such case   solutions to the Fourier equation approximate well asymptotic solutions to the Cattaneo equations.

In order to analyze asymptotic solutions to (\ref{Catt}) we note its formal similarity to the Hamilton equations in particle mechanics. The first equation corresponds to $\dot{r}=p^*$, and the second to $\dot{p}= -r^*-\lambda p^*$, where $r$ is the position vector, $p$ the momentum, $(r^*,p^*)$ are the conjugate variables, and $\lambda p^*$, $\lambda>0$ represents the friction. In this context, we know that if the mass $m$ of the particle (that is involved in the relation between $p$ and $p^*$) is sufficiently small and  $\lambda$ (that is involved in the friction)  is sufficiently large then the $p$ evolves  faster than $r$. After some time, $p$ settles at (or at the vicinity of) the quasi-equilibrium  manifold determined by $r^*-\lambda p^*=0$. The time evolution then continues on the quasi-equilibrium manifold and is governed by the inertialess dynamics $\dot{r}=-\frac{1}{\lambda}r^*$.

Analogically in the context of (\ref{Catt}), with an appropriate choice of $s^{(M)}$ (determining the relation between the variables and their conjugates) and $\Psi^{(M)}$  (determining the dissipation), $\JJ$  evolves  in time faster than $e$ and consequently, at the later stage of the time evolution,  both $\partial\JJ/\partial t$ and $\JJ$ are   small so that the second equation in (\ref{Catt}) reduces to
\begin{equation}\label{qsm}
\frac{1}{(e^*)^2}\partial_i(e^*)+\Psi^{(M)}_{J^*_i}=0
\end{equation}
We note that with  the same choice of $s^{(M)}$ and  $\Psi^{(M)}$  and at the same stage of the time evolution,  the second term on the right hand side of (\ref{PBph}) will be  smaller than the first term. 
Indeed, the governing equations implied by Poisson bracket \eqref{PBph} are
\begin{equation*}
\frac{\partial}{\partial t}\left(\begin{array}{cc}e\\J_i\end{array}\right)=\left(\begin{array}{cc}\partial_i\left(\frac{1}{(e^*)^2}J^*_i\right)\\\frac{1}{(e^*)^2}\partial_i(e^*)\end{array}\right)-\boxed{\frac{1}{s^{(M)}} \left(\partial_i J_j-\partial_j J_i\right)\begin{pmatrix} \left(\frac{J_i^*}{e^*}\right)^2 \\ \frac{J_i^*}{e^*}\end{pmatrix}}
+\left(\begin{array}{cc}0\\\Psi^{(M)}_{J^*_i}\end{array}\right)
\end{equation*}
where the boxed terms are contributions from the second term of the Poisson bracket. One can see that near the Fourier manifold, where $J_i\ll 1$ these terms are subleading to the other terms justifying their omission above. In the one-dimensional setting, moreover, the terms are equal to zero exactly. Similarly, time evolution of the flux $\JJ$ near the Fourier manifold is well approximated by the algebraic constraint \eqref{qsm}.

Summing up, the above considerations  imply that (with an appropriate choice of $s^{(M)}$ and  $\Psi^{(M)}$ and limiting ourselves  only the later stage of the time evolution) the time evolution
governed by (\ref{Catt})  proceeds in two stages.

In the first (fast) stage the state variables approach the vicinity of the quasi-equilibrium manifold  determined by (\ref{qsm}). The time evolution in this stage (describing the approach $M\rightarrow m$) is governed by (\ref{Ups})  with $\Upsilon=\JJ^*$,  the Constitutive Relation $\mathcal{J}^{CR}_i=-\frac{1}{(e^*)^2}\partial_i e^*$, and
\begin{equation}\label{PhiPsi}
S^{(Mm)}=\Psi^{(M)}.
\end{equation}
Indeed, the fast (reducing) evolution equation then reads
\begin{equation}
	\dot{J}^*_i =  \frac{\partial \delta s^{(M)}/\delta J_i}{\partial t} = \frac{\delta^2 s}{\delta J_i \delta J_j} \dot{J}_j =  s^{(M)}_{J_i J_j} \frac{\delta}{\delta J_j} \left(\Psi^{(M)} + \int \dr \left(\frac{1}{(e^*)^2}\partial_i e^*\right) J^*_i\right)
	=
      \end{equation}
while $\dot{J}^*_i = \dot{\Upsilon}_i =  -\Lambda^{(Mm)}_{ij} \Phi^{(Mm)}_{J^*_j}$ and where $\Lambda^{(Mm)}_{ij} = -s^{(M)}_{J_i J_j}$ is a symmetric positive definite operator (due to concavity of entropy). Therefore 
\begin{equation}
	\Phi^{(Mm)} = -\Psi^{(M)} + \int \dr \left(-\frac{1}{(e^*)^2}\partial_i e^*\right) J^*_i
\end{equation}
plays the role of thermodynamic potential.

In the second stage the time evolution is governed by (\ref{Four2}) with
\begin{equation}\label{PsimM}
\Psi^{(m)}=\left[\Psi^{*(M)}(X^{(M)})\right]_{X^{(M)}_i=-\frac{1}{(e^*)^2}\partial_ie^*},
\end{equation}
where $\Psi^{*(M)}$ is the dissipation potential conjugate to $\Psi^{(M)}$ via Legendre transformation
\begin{subequations}
\begin{equation}
	0 = \frac{\delta}{\delta \JJ^*}\left( - \Psi^{(M)}+ \XX^{(M)} \cdot \JJ^*\right) 
	\qquad\mbox{and}\qquad \Psi^{*(M)} = -\Psi^{(M)}\left(\JJ^*\left(\XX^{(M)}\right)\right) + \XX^{(M)}\cdot\JJ^*\left(\XX^{(M)}\right).
\end{equation}
The inverse transformation reads
\begin{equation}
	0 = \frac{\delta}{\delta \XX^{(M)}}\left( - \Psi^{*(M)}+ \XX^{(M)} \cdot \JJ^*\right) 
	\qquad\mbox{and}\qquad \Psi^{(M)} = -\Psi^{*(M)}\left(\XX^{(M)}(\JJ^*)\right) + \XX^{(M)}(\JJ^*)\cdot\JJ^*.
\end{equation}
\end{subequations}
For instance, if $\Psi^{(M)} = \int \dr \frac{1}{2}\lambda \JJ^*\cdot\JJ^*$, then $\Psi^{*(M)}= \int \dr \frac{1}{2}\frac{1}{\lambda}\XX^{(M)}\cdot\XX^{(M)}$ and 
\begin{equation}
	\Psi^{(m)} = \int \dr \frac{1}{2\lambda (e^*)^4}(\nabla e^*)^2.
\end{equation}
Indeed, solutions to (\ref{qsm}) are
\begin{equation} \label{34}
	J^*_i=\left[\Psi^{*(M)}_{X^{(M)}_i}\right]_{X^{(M)}_i=-\frac{1}{(e^*)^2}\partial_i e^*}.
\end{equation}
Evolution equation for energy density in Eq. \eqref{Catt} becomes 
\begin{equation}
	\partial_t e = \partial_i \left(\frac{1}{(e^*)^2} \Psi^{*(M)}_{X^{(M)}_i}\Big|_{\XX^{(M)} = \nabla \frac{1}{e^*}}\right),
      \end{equation}
and hence energy conservation is guaranteed.

Particularly interesting and important is the finding that the Mm-entropy $S^{(Mm)}$  that arises in the analysis of the passage $M\rightarrow m$, becomes directly related to the ``$m$'' dissipation potential $\Psi^{(m)}$ that arises in the analysis of the passage $m\rightarrow ET$. The relation is given by (\ref{PhiPsi}) and (\ref{PsimM}).  In other words, in systems that are not prevented from reaching equilibrium, the entropy that arises in the investigation of the passage $M\rightarrow m$ is a quantity directly related to the production of the entropy that arises in the investigation of the passage $m\rightarrow ET$.

In particular, in the situation at which the second equation in \eqref{Catt} reduces to its equilibrium form \eqref{qsm} we also replace the $s_e^{(M)}$ with $s_e^{(m)}$ (i.e. $e^*$ on the ``M'' level is the same as $e^*$ on the ``m'' level) in accordance with MaxEnt reduction \cite{RedExt}. 
From \eqref{34} we then get
\begin{equation*}
  \dot{S}^{(m)} = \langle \Upsilon, \XX \rangle = \int \dr \left[ X_i^{(M)} \Psi^{*(M)}_{X_i^{(M)}}\right]_{\XX^{(M)}=\nabla\frac{1}{e^*}}>0
\end{equation*}
and hence the entropy inequality is satisfied. Only the entropy production on m-level is the above expression instead of $\int \dr e^* \Psi_{e^*}^{(m)}$ both of which are positive. As a result the relation between the Mm-entropy and ``m'' dissipation potential is more complicated.

\section{Extended Cattaneo level}
It was shown for instance in \cite{KovacsVan-BalisticHeat18} that good agreement with experiments is obtained when working with not only the $\JJ$ field (Cattaneo $M$ level), but also with an additional tensor field $\QQ$ coupled to the $\JJ$ field. Let us now show how to derive the equations using CR-thermodynamics.

A natural CR-extension of the Cattaneo $M$-level is the $EC$ (extended Cattaneo) level with state variables $(s, \JJ, \QQ)$. The Poisson bracket is an extension of bracket \eqref{PBph} (disregarding the higher order terms in that bracket as above),
\begin{equation}\label{eq.PB.EC}
	\{A,B\}= \int \dr \left(\partial_k A_{s^{(M)}} B_{J_k}-\partial_k B_{s^{(M)}} A_{J_k}\right)
	+\int \dr \left(\partial_j A_{J_i} B_{Q_{ij}}-\partial_j B_{J_i} A_{Q_{ij}}\right).
\end{equation}
The reversible evolution equations implied by Poisson bracket \eqref{eq.PB.EC} are
\begin{subequations}
	\begin{eqnarray}
		\partial_t s &=& -\partial_i E_{J_i}\\
		\partial_t J_i &=& -\partial_i E_s -\partial_j E_{Q_{ij}}\\
		\partial_t Q_{ij} &=& -\partial_j E_{J_i}.
	\end{eqnarray}
\end{subequations}
Transformation from this energetic representation to the entropic representation turns the evolution equations to
\begin{subequations}
	\begin{eqnarray}
	\frac{\partial e}{\partial t}&=&\partial_i\left(\frac{1}{(e^*)^2}J^*_i\right)\\
		\frac{\partial J_i}{\partial t}&=& \frac{1}{(e^*)^2}\partial_i(e^*) + \partial_j \frac{Q^*_{ij}}{e^*}\\
		\frac{\partial Q_{ij}}{\partial t}&=& \partial_j \frac{J^*_i}{e^*},
	\end{eqnarray}
\end{subequations}
where the conjugate variables $e^*$, $\JJ^*$ and $\QQ^*$ can be interpreted as derivatives of a $S^{(EC)}$ entropy living on the $EC$ level.
This is the reversible evolution of state variables $(e, \JJ, \QQ)$. 

Dissipation is included through a dissipation potential on the $EC$ level $\Psi^{(EC)}(\JJ^*,\QQ^*)$, and the evolution equations become
\begin{subequations}\label{eq.evo.EC}
	\begin{eqnarray}
	\frac{\partial e}{\partial t}&=&\partial_i\left(\frac{1}{(e^*)^2}J^*_i\right)\\
		\frac{\partial J_i}{\partial t}&=& \frac{1}{(e^*)^2}\partial_i(e^*) + \partial_j \frac{Q^*_{ij}}{e^*}+ \Psi^{(EC)}_{J^*_i}\\
		\frac{\partial Q_{ij}}{\partial t}&=& \partial_j \frac{J^*_i}{e^*}+ \Psi^{(EC)}_{Q^*_{ij}}.
	\end{eqnarray}
\end{subequations}
The choice 
\begin{equation}
	\Psi^{(EC)}(\JJ^*,\QQ^*) = \int\dr \frac{1}{2}\lambda (\JJ^*)^2 + \int\dr \frac{1}{2}\alpha Q^*_{ij} Q^*_{ij}
\end{equation}
makes the evolution equations explicit (up to the specification of entropy),
\begin{subequations}\label{eq.evo.EC.exp}
	\begin{eqnarray}
	\frac{\partial e}{\partial t}&=&\partial_i\left(\frac{1}{(e^*)^2}J^*_i\right)\\
		\frac{\partial J_i}{\partial t}&=& \frac{1}{(e^*)^2}\partial_i(e^*) + \partial_j \frac{Q^*_{ij}}{e^*}+ \lambda J^*_i\\
		\frac{\partial Q_{ij}}{\partial t}&=& \partial_j \frac{J^*_i}{e^*}+\alpha Q^*_{ij}. 
	\end{eqnarray}
\end{subequations}
These equations are compatible with equations (16) of paper \cite{KovacsVan-BalisticHeat18}.

The evolution equation for the $\QQ$ field in Eqs. \eqref{eq.evo.EC.exp} can be seen as the reducing evolution when assuming fast relaxation of $\QQ$ and leading (after the relaxation) to equation
\begin{equation}
		0 = \partial_j \frac{J^*_i}{e^*}+\alpha Q^*_{ij},
\end{equation}
which is a constitutive relation for $\QQ^*$ in the evolution equation for $\JJ$. After plugging this constitutive relation into the evolution equation for $\JJ$ we obtain

\begin{subequations}
	\begin{eqnarray}
		\frac{\partial e}{\partial t}&=&\partial_i\left(\frac{1}{(e^*)^2}J^*_i\right)\\
		\frac{\partial J_i}{\partial t}&=& \frac{1}{(e^*)^2}\partial_i(e^*) 
		- \partial_j \left(\frac{1}{\alpha e^*}\partial_j \frac{J^*_i}{e^*}\right)+ \lambda J^*_i,
	\end{eqnarray}
\end{subequations}
which can be interpreted as the Guyer-Krumhansl equations, see e.g. \cite{Kovacs-Van-dual,Kovacs-Van15}. 

In summary, an another extension of the Cattaneo level by adding an extra tensor field $\QQ$ leads to a nonlinear generalization of equations that have been shown in good agreement with flash experimental data. By reduction of the fast evolution we obtain a generalization of the Guyer-Krumhansl equations.

\section{Discussion}

Emergence of various entropies in the analysis of the time evolution of both externally unforced and driven systems has already  been discussed in \cite{Grjsf}, \cite{Grrew}. In this paper we have  worked out a simple illustration. A systematic investigation of relations among the heat conduction theories formulated on three levels, namely the  equilibrium ``$ET$'', the Fourier ``$m$'', and the Cattaneo ``$M$'' levels,  led us to seven  entropies, three on ``$ET$'', two on ``$m$'' and  two on ``$M$'' levels.
\\

\textbf{\textit{ET-entropies}}

The first ET-entropy, introduced in (\ref{fundET}), arises as a result of experimental observations (listed in Thermodynamic Tables). The second and the third ET-entropies arise in Sections \ref{Ftheory} and \ref{Cattth} in the analysis of solutions to the Fourier and the Cattaneo heat conduction equations.  Their origin is thus dynamical. It is the process that is needed to prepare the systems to the ET-level that is giving rise to the ET-entropies.
\\

\textbf{\textit{m-entropies; the main result}}

The first m-entropy arises in Section \ref{Ftheory} as a potential generating the preparation process for using the equilibrium level of description. This entropy therefore does not exist if the system under consideration is prevented from reaching the thermodynamic equilibrium.
The second m-entropy, that arises in Section \ref{4} in the analysis of the passage from the Cattaneo to the Fourier description of the heat conduction, however exists also in the presence of external forces and external and internal constraints preventing the approach to the thermodynamic equilibrium. \textbf{The second m-entropy thus provides thermodynamics also to  systems for which the classical equilibrium thermodynamics does not exist.
If the approach to equilibrium is permitted then both m-entropies exist and the latter (i,e, the one associated with the passage $M\rightarrow m$) turns out to be   production of the former (i,e, the one associated with the passage $m\rightarrow ET$)}. The m-entropy that arises in the passage $M\rightarrow m$ can be therefore called an entropy production but such terminology is confusing since such entropy production does not have to be production of any entropy (as it is indeed  the case when the  approach to equilibrium does not exist and thus the classical entropy does not exist). We therefore suggest to call the entropy that arises in the analysis of $M\rightarrow m$ a CR-entropy, i.e. the entropy determining the Constitutive Relations. The  realization that  the classical entropies and the CR-entropies have  very different origins brings also a clarification to discussions about the maximum-entropy and the maximum-entropy-production principles (for example in their use in determining the constitutive relations in fluid mechanics of complex fluids \cite{Raj}, \cite{PepaRaj}).
\\

\textbf{\textit{M-entropies}}

There are two M-entropies, one arising in the analysis of $M\rightarrow ET$ and the other in the analysis of $M\rightarrow m$. We note that the transformation of  the M-entropy corresponding to the passage $M\rightarrow ET$ to  the resulting ET-entropy, that is made by following the time evolution on the level ``$M$'' to its conclusion is also a reducing Legendre transformation. This is true also for the transformation from the m-entropy to the corresponding to it ET-entropy in the analysis of $m\rightarrow ET$ and the transformation from the Mm-entropy to the m-entropy in the passage $M\rightarrow m$.
\\
\\

\section*{Acknowledgements}

This research has been supported partially by the Natural Sciences and Engineering Research Council of Canada, Grants 3100319 and 3100735.
This work was also supported by Czech Science Foundation, project no.  17-15498Y, and by Charles University Research program No. UNCE/SCI/023. 
We are grateful to P{\' e}ter V{\' a}n, Tam{\' a}s F{\" u}l{\" o}p, R{\' o}bert Kov{\' a}cs and M{\' a}ty{\' a}s Sz{\" u}cs for pleasant and fruitful discussions during our Budapest meeting, that motivated some parts of this paper.


\begin{thebibliography}{}

\bibitem{KovacsVan-BalisticHeat18}

  R. Kov{\'a}cs and P. V{\'a}n, Second sound and ballistic heat conduction: NaF experiments revisited, Int J Heat and Mass Transf., 117 (2018), 682--690.



	
\bibitem{Cattaneo}

 C. Cattaneo, Sulla conduzione del calore, Atti Seminario Mat. Fis. Univ. Modena. 3 (1948), 83–101.


\bibitem{Vill1}
L. Desvillettes and C. Villani, On the trend to global equilibrium for spatially inhomogeneous
kinetic systems: The Boltzmann equation, Invent. Math., 159 (2005), 245-316.

\bibitem{GO}
M. Grmela, H.C. \"{O}ttinger,  Dynamics and thermodynamics of
complex fluids. I. Development of a general formalism. Phys.
Rev. E 56 (1997) 6620-6632



\bibitem{OG}
H.C. \"{O}ttinger,  M. Grmela,  Dynamics and thermodynamics of
complex fluids. II. Illustrations of a general formalism. Phys.
Rev. E 56 (1997) 6633-6655







\bibitem{O}
H.C. \"{O}ttinger,  Beyond Equilibrium Thermodynamics; Wiley: New York, NY, USA, (2005).




\bibitem{PKG}
M. Pavelka, V. Klika, M. Grmela, Multiscale Thermo-Dynamics (2018) De Gruyter

\bibitem{Temam}
  R. Temam, Infinite-dimensional dynamical systems in mechanics and physics (2012); Springer Science \& Business Media.
  

\bibitem{GT}
M. Grmela and J. Teichmann, Lagrangian formulation of Maxwell-Cattaneo hydrodynamics, Int. J. Eng. Sci. 21 (1983),



\bibitem{Guo1}
Y. Dong, B.-Y. Cao and Z.-Y. Guo, Generalized heat conduction laws based on thermomass theory and phonon hydrodynamics,
J. Appl. Phys. 110 (2011), 063504.



\bibitem{Vitek}
Miroslav Bul{\' i}{\v c}ek, Josef M{\' a}lek and V{\' i}t Pr{\r u}{\v s}a, Thermodynamics and stability of non-equilibrium steady states in open systems, arXiv cond-mat.stat-mech: 1709.05968, 2018..





\bibitem{RA}
T. Ruggeri, M. Sugiyama,  Rational Extended Thermodynamics Beyond the Monoatomic Gas. Springer,
Heidelberg (2015)







\bibitem{Arnold1}
V. I. Arnold, Sur la g\'{e}ometrie diff\'{e}rentielle des groupes
de Lie
de dimension infini et ses applications dans l'hydrodynamique
des fluides parfaits, Ann. Inst. Fourier 16,  (1966) 319






\bibitem{Raj}
 K.{\,}R. Rajagopal and A.{\,}R. Srinivasa,
 A thermodynamic framework for rate type fluid models,
 J. Non-Newtonian Fluid Mech. 88, (2000) 207-227



\bibitem{PepaRaj}
  J. M\'{a}lek and K.{\,}R. Rajagopal and K. Tůma,
  On the variant of the {M}axwell and {O}ldroyd-{B} models with the context of a thermodynamic basis,
  Int. J. Non-Linear Mech. 76, (2015) 42-47




\bibitem{Grjsf}
M. Grmela, Externally driven macroscopic systems: Dynamics
versus thermodynamics, J. Stat. Phys. 166, (2017) 282

\bibitem{Buis}
G. R. Buis,  W. G. Vogt and M. M. Eisen,
Lyapunov stability for partial diferential equations,
Dissertation at University of Pittsburgh, 1968


\bibitem{Grrew}
M. Grmela. Generic guide to the multiscale dynamics and thermodynamics. Journal of
Physics Communications, 2(032001), 2018.

\bibitem{RedExt}
Grmela, M., Klika, V. and Pavelka, M., Reductions and Extensions in mesoscopic dynamics, Phys. Rev. E 92 (2015), 032111.

\bibitem{Kovacs-Van-dual}
R. Kov{\'a}cs and P. V{\'a}n, Thermodynamical consistency of the dual-phase-lag heat conduction equation, 
Continuum Mechanics and Thermodynamics, 2017 (accepted), DOI:10.1007/s00161-017-0610-x.

\bibitem{Kovacs-Van15}
R. Kov{\' a}cs and P. V{\' a}n, Generalized heat conduction in heat pulse experiments, 	
International Journal of Heat and Mass Transfer 83 (2015) 613-620

\bibitem{ADER1}
Michael Dumbser, Ilya Peshkov, Evgeniy Romenski, Olindo Zanotti, High order ADER schemes for a unified first order hyperbolic formulation of continuum mechanics: Viscous heat-conducting fluids and elastic solids,
Journal of Computational Physics 314 (2016), 824-862.

\bibitem{SHTC-GENERIC}
Ilya Peshkov, Michal Pavelka, Evgeniy Romenski, Miroslav Grmela, Continuum Mechanics and Thermodynamics in the Hamilton and the Godunov-type Formulations, Accepted to Continuum Mechaanics and Thermodynamics 2018.

\end{thebibliography}
\end{document}